*The role of chemo-mechanical modelling in the development of battery technology – a perspective*

Adam M. Boyce[1], Emilio Martínez-Pañeda[2], Paul R. Shearing[2,3,*]

[1]School of Mechanical and Materials Engineering, University College Dublin, Dublin D04V1W8, Ireland
[2]Department of Engineering Science, University of Oxford, Oxford OX1 3PJ, United Kingdom
[3]The Faraday Institution, Quad One, Didcot OX11 0RA, United Kingdom
*Corresponding author: paul.shearing@eng.ox.ac.uk

In the race to reduce global CO2 emissions and achieve net-zero, chemomechanics must play a critical role in the technological development of current and next-generation batteries to improve their energy storage capabilities and their lifetime. Many degradation processes arise through mechanics via the development of diffusion-induced stress and volumetric strains within the various constituent materials in a battery. From particle cracking in lithium-ion batteries to lithium dendrite-based fracture of solid electrolytes in solid-state batteries, it is clear that significant barriers exist in the development of these energy storage systems, where chemomechanics plays a central part. To accelerate technological and scientific advances in this area, multi-scale and highly coupled multiphysics modelling must be carried out that includes mechanics-based phenomena. In this perspective article, we provide an introduction to chemomechanical modelling, the various physical problems that it addresses, and the issues that need to be resolved in order to expand its use within the field of battery technology.

**Introduction**

The operating mechanisms of a battery are exceedingly complex, involving multiple, highly coupled physical processes that include electrochemistry, species transport (such as diffusion, migration and convection), as well as thermodynamics and solid mechanics. Consideration of solid mechanics has become increasingly pertinent since material stress and strain play a crucial role in the performance and lifetime of not only the ubiquitous lithium-ion battery (LIB), but also next-generation technologies such as solid-state (SSB) and sodium-ion batteries (SIB).

While computational modelling of electrochemical, transport and thermal processes for battery applications has received much attention, the consideration of solid mechanics, or chemomechanics, is in its infancy; this is partly due to the computational expense of these simulations. In the race to produce energy storage solutions with high energy density and long lifetimes, there exists some key barriers that are based on chemomechanics and we highlight two examples as follows: 1) diffusion-induced stress and volumetric straining of the active materials due to lithium insertion or extraction in cathodes and anodes of LIBs, SSBs and SIBs. 2) the initiation and growth of voids and lithium dendrites in lithium anodes of SSBs via plating and stripping [1–4]. These two examples, as seen in Figure 1(a)-(d), are typically associated with the fracture of various components and subsequently lead to degradation, short-circuiting and general failure of batteries.

To improve our understanding of performance, degradation processes and safety, it is crucial that we develop a full picture of their physical behaviour using advanced, physically grounded chemomechanical models. In this article, we give a brief overview and introduction to the role that chemomechanical modelling plays in battery research as well as a perspective on the future of this line of investigation. The interested reader can find in-depth and comprehensive reviews on the broader subject in the works of Deshpande and McMeeking [5] and Zhao et al. [6].

**Detailed insight into the role of mechanics and how it's modelled**

When an ionic species is reduced or oxidised, lithium (or sodium) is either inserted or extracted from the host active material, depending on the reaction direction. These changes in the crystal structure of the material cause lattice strain. The lithiation induced strain, $\varepsilon$, is typically given by the following expression

$$\varepsilon_L = \frac{1}{3}\Omega(c - c_0)\mathbf{I}$$

Where $\Omega$ is the partial molar volume, $c$ is the lithium concentration, $c_0$ is the initial lithium concentration and $\mathbf{I}$ is the identity tensor. In terms of thermodynamics, the total free energy in an active material is a function of the electrochemical energy and the strain energy that exists when lithium is inserted or alloyed [7]. The related chemical potential $\mu$ is as follows

$$\mu = \mu_0 + RT \ln\left(\frac{c}{c - c_{max}}\right) - \Omega\sigma_H$$

Where, $\mu_0$ is a reference chemical potential, $R$ is the universal gas constant, $T$ is the temperature, $c_{max}$ is the maximum lithium concentration and $\sigma_H$ is the hydrostatic stress. Consequently, diffusive flux, $J$, of lithium in the active materials is also affected by the hydrostatic stress

$$J = -D\left(\nabla c - \frac{\Omega c}{RT}\nabla \sigma_H\right)$$

Where $D$ is the diffusion coefficient. Thus, if diffusion-induced stresses and volumetric straining occurs during operation, battery capacity can be directly affected by increasing or decreasing the overpotentials, since overpotential is related directly to the electrochemical potential. Thus, there exists a complex coupling between solid mechanics, transport of lithium-species and the kinetics of the electrochemical reaction. In the case of intercalation-type electrode materials such as graphite and transition layered oxides, e.g., NMC, LCO, LMO, strains are typically on the order of 5% and contribute a modest amount to the total energy. However, when materials such as lithium, sodium and silicon or magnesium-lithium alloys are considered, then strains up to 300% may be attained [8,9], giving rise to a significant change in stored energy and general battery performance [10–13]. These mechanical and thermodynamic considerations form the basis for all chemomechanical models for batteries.

Here, we have given a very brief summary of the primary equations that link electrochemistry and mechanics – detailed derivations and full formulations can be found in [5,12,14].

As with all theoretical and computational frameworks, an appropriate constitutive formulation must be chosen to serve as the underlying foundation for a given chemomechanical model. In the case of transition metal layered oxides, it is typical to consider a linear elastic Hooke's law since the strains are low. In the case of secondary agglomerates, this may be acceptable given that fracture of the particle occurs before any plasticity develops. However, when considering a single crystal of active material, slip may occur prior to fracture and there is ongoing research in this area [15]. In metallic active materials such as lithium or silicon, large-scale rate-dependent deformation occurs and viscoplasticity is typically considered [12]. This adds significant complexity such that strain rate and consequently battery charge and discharge rates dramatically change the mechanical and electrochemical performance of the battery. Batteries in general are multi-material composites and deformation of all associated materials must be considered. The active materials typically swell or contract upon insertion or removal of lithium and will impart a load upon the adjacent materials such as the carbon binder in liquid-based LIBs. It is thus imperative that the surrounding materials are adequately captured, both in terms of constitutive models and in terms of boundary conditions, e.g. interfacial contact and constraint. Furthermore, there is a significant difference between the deformation response of a liquid-based LIB and SIB, and a solid-based SSB and SIB. The electrolytes in SSBs will provide much greater constraint and this is an area where chemomechanics plays a crucial role, since loss of contact due to interfacial fracture at the particle-solid electrolyte is detrimental to battery performance [16,17].

Fracture plays a central role in the long-term stability of battery materials, and it is crucial that it is considered from a modelling perspective [18]. Consider the layered oxide materials that are used as a cathode in an LIB, which are typically polycrystalline (i.e. formed as agglomerates of primary particles) or single crystal. A given crystal of these materials strains in an anisotropic fashion when lithium is inserted due to their layered nature [15,19]. Once these randomly oriented single crystals (primary particles) are assembled/bonded together in their thousands, bulk fracture of the so-called agglomerate secondary particle occurs when (de)lithiation occurs, see Figure 1(c). This results in the inflow of electrolyte, resulting in build-up of parasitic interphases that reduce battery lifetime, while also resulting in the detachment of particles from the surrounding carbon additives that render them electrically isolated and inactive. Fracture also acts as a barrier to the adoption of SSBs – loss of contact between the solid electrolyte and cathode or anode active materials due to particle swelling leads to poor battery lifetime. Battery research and hence chemomechanical modelling is a multiscale problem and deformation and fracture occurs at the atomic level [20] and particle level as described, whilst also occurring at the electrode and cell scales. At the electrode-scale, transport and kinetics limitations, typical of moderately thick electrodes, give rise to lithium concentration gradients within the particles, which in turn results in stress gradients and subsequent fracture in a heterogeneous fashion [21]. To enhance the energy density of lithium-ion batteries, thick electrodes are required and chemomechanics thus plays an important role. Furthermore, batteries are assembled into many layers of electrodes at the cell-level, leading to a significant build-up of in-homogeneous deformation due to heterogeneous phenomena at the aforementioned particle and electrode levels [22]. Failure of the batteries at the cell-level can be detrimental and extremely hazardous, where the safe use at the end application is concerned, such as an electric vehicle. These examples highlight the need for appropriate fracture modelling within the finite element, or finite volume environment – the typical approach for computational models. Cohesive zone elements that predict crack growth along a known and predefined path are typically used [23], while more recently, the X-FEM [24,25] and phase field methods have been employed to predict complex crack propagation in arbitrary directions and without requiring prior knowledge of the crack location [21,26–28]. Furthermore, molecular dynamics simulations have been used to model fracture at a more fundamental atomistic level [20].

Material properties and parameters play a critical role in any model and govern the accurate representation of the predicted mechanism in question - this is no different in the case of chemomechanics. In contrast to regular solid mechanics problems, properties such as Young's modulus, Poisson's ratio, and yield strength may change as a function of state-of-lithiation, i.e. the lithium concentration level in an active material, adding further complexity to these multiphysics models [29]. Furthermore, some active materials such as silicon, and lithium iron phosphate undergo phase changes during lithium insertion or extraction [30]. This behaviour must also be accounted for in computational models since changes in properties may occur during phase change and can change battery performance as a result. Models and experiments must act in tandem; accurate property measurement must be achieved in order to facilitate chemomechanical models. Numerous challenges exist on this front since batteries are a multiscale problem and most of the material properties in question must be measured at the micron or nano scale. For example, most plasticity models are typically parameterised through the material stress-strain curve response, measured using a tensile test, which requires large samples that are very difficult to obtain given the scales in question. Measurements may be further hindered, if concentration-dependency of material properties is considered. Finally, as with all experiments, repeatability must be achieved in order to facilitate high-fidelity models. This is necessitated by the highly heterogeneous nature of batteries at all scales - accurate and repeatable parameters must be measured and then, crucially, rigorous experimental validation of the model must be achieved.

As mentioned, battery research is a multiscale problem and thus battery models must be formulated as such with the necessary linking required. Of most interest are the single particle (SPM) and electrode scales [31]. Numerical continuum models such as these are usually solved using the finite element or finite volume method with suitable discretised partial differential equations and geometrical mesh. Most SPM works have considered the agglomerated secondary particle as a homogenised entity, without resolving the constituent primary particles - this provides great insight into the influence of void-based, concentration gradient-driven fracture [26,27,30,32–34]. The aforementioned and additional mechanism of anisotropy-driven cracking when the primary particles are resolved has also been considered [28,35]. Fatigue crack growth is also an important consideration given the cyclic nature of battery operation [36]. Electrode-scale models, usually based on X-ray computed tomography images, are concerned with the same mechanisms at a coarser resolution but permit detailed insights across the entire electrode microstructure, allowing deformation and cracking of multiple phases such as carbon binder, solid electrolyte and multiple particles, see Figure 1(d) [14,21,37]. They also enable the actual geometry to be captured within the model, allowing heterogeneities in the strain field to be adequately resolved. While the electrode-scale models are highly valuable from a scientific perspective, they are computationally expensive and rely on costly imaging equipment – these issues must be addressed to develop improved chemomechanical models. To make these electrode-scale chemomechanical models industrially applicable, e.g. in battery management systems, work must be carried out to implement their findings into models that are more suited to rapid electrode or cell design such as the Doyle-Fuller-Newman (DFN) model [31].

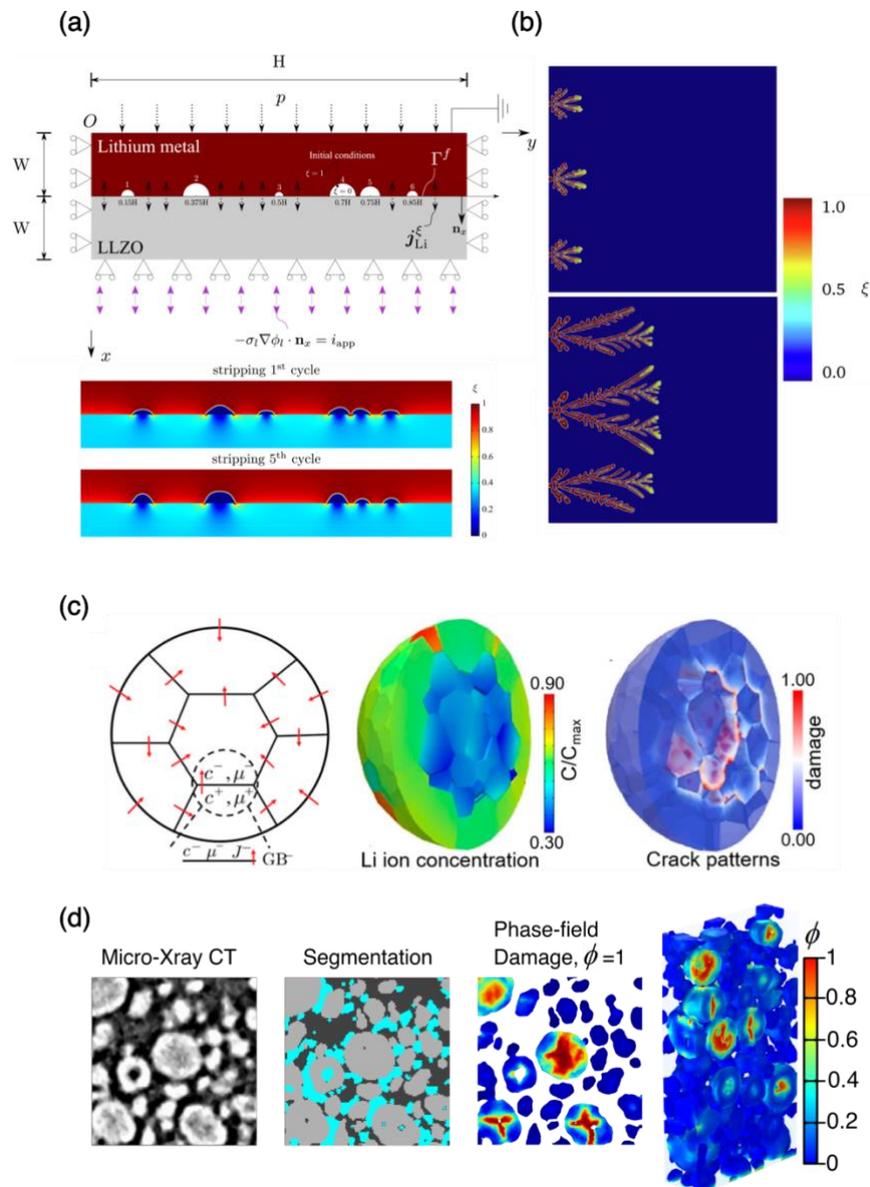

**Figure 1.** (a) Phase field model for void evolution in lithium-metal anodes for SSBs. Reprinted (adapted) with permission from [3]. Copyright 2022 Elsevier. (b) Predictions of dendrite growth in a solid electrolyte using a phase field model. Reprinted (adapted) with permission from [4]. Copyright 2022 Elsevier. (c) Cohesive zone element-based fracture model for anisotropy driven fracture in LIBs. Reprinted (adapted) with permission from [38]. Copyright 2022 American Chemical Society. (d) Concentration gradient-driven phase field fracture model for LIBs Reprinted (adapted) with permission from [21]. Copyright 2022 Elsevier.

**Outlook**

Chemomechanics is a burgeoning and critical area of research in energy storage and battery technology. It is clearly of ongoing interest in the LIB and SSB communities, and a necessity for upcoming technologies such as the SIB . To solve the problems highlighted throughout this article it is worth noting that battery technology sits on a multiphysics foundation and continued and sustained interdisciplinary collaboration will be required – this should include mechanical engineers, physicists, and materials scientists to overcome the various prohibitive chemomechanical issues that prevail. Technological development of battery systems requires the intersection of computational modelling and experiment; experiments fulfil the crucial role of physical insight, while also providing the basis for computational models that provide a cost-effective way of design and optimisation as well as enabling further physical insight. Engineering and scientific experimentalists must therefore be enlisted to support chemomechanical modelling efforts. Chemomechanical modelling has shown great importance when applied to conventional LIBs. While it is still crucial that this research continues in the suggested areas, the increasing focus on lithium or sodium-based solid-state batteries requires much attention where chemomechanics is concerned. Given the issues related to pulverisation of alloy-type anodes due to large volume changes, detrimental short circuiting due to dendrite growth in the case of lithium anodes, and interfacial contact loss between solid electrolyte and cathode materials, chemomechanical modelling and experiments must be front and centre in efforts to enhance the performance of next generation battery technology.

**Acknowledgements**

Emilio Martínez-Pañeda was supported by an UKRI's Future Leaders Fellowship [grant MR/V024124/1]. Paul Shearing would like to acknowledge the Royal Academy of Engineering (CiET1718\59) for financial support and the Faraday Institution (faraday.ac.uk; EP/S003053/1), grant number FIRG015.